\PassOptionsToPackage{dvipsnames}{xcolor}

\documentclass[sigconf,nonacm]{acmart}
\usepackage{mathtools}
\usepackage{array}
\usepackage{enumitem}
\usepackage{tikz}
\usetikzlibrary{arrows.meta,positioning,shapes.geometric,shapes.multipart,fit,backgrounds,calc}
\usepackage{algorithm}
\usepackage{algpseudocode}

\definecolor{navy}{RGB}{16,27,51}
\definecolor{teal}{RGB}{14,124,134}
\definecolor{amber}{RGB}{217,137,42}
\definecolor{slate}{RGB}{91,107,126}
\definecolor{lightblue}{RGB}{214,236,236}
\definecolor{geneblue}{RGB}{52,101,164}
\definecolor{pathgreen}{RGB}{30,130,76}
\definecolor{disred}{RGB}{175,54,42}
\definecolor{passgray}{RGB}{140,140,140}

\theoremstyle{acmdefinition}
\newtheorem{definition}{Definition}[section]
\newtheorem{example}{Example}[section]
\newtheorem{problem}{Problem}[section]
\newtheorem{remark}{Remark}[section]
\theoremstyle{acmplain}
\newtheorem{proposition}{Proposition}[section]
\newtheorem{theorem}{Theorem}[section]

\DeclareMathOperator{\rob}{rob}

\newcommand{\KG}{\mathcal{G}}
\newcommand{\LG}{\mathcal{L}_{\KG}}
\newcommand{\logicname}{BKML}

\newcommand{\bools}{\{0,1\}}
\newcommand{\R}{\mathbb{R}}

\newcommand{\VGene}{V_{\mathrm{Gene}}}
\newcommand{\AP}{\mathit{AP}}
\newcommand{\Ops}{\mathbf{Op}}

\tikzset{
  gene/.style={circle, draw=geneblue, fill=geneblue!12, thick,
               minimum size=1.8cm, font=\small\bfseries\color{geneblue},
               text width=1.5cm, align=center, inner sep=1pt},
  passenger/.style={circle, draw=passgray, fill=passgray!12, thick,
               minimum size=1.8cm, font=\small\bfseries\color{passgray},
               text width=1.5cm, align=center, inner sep=1pt},
  pathway/.style={rectangle, rounded corners=3pt, draw=pathgreen, fill=pathgreen!12,
               thick, minimum height=0.75cm, minimum width=2.0cm,
               font=\small\bfseries\color{pathgreen}, align=center},
  disease/.style={diamond, draw=disred, fill=disred!10, thick,
               minimum size=1.4cm, font=\small\bfseries\color{disred}, align=center,
               inner sep=0pt, aspect=2.2},
  assocwith/.style={-{Stealth[length=6pt]}, thick, color=disred,
               dashed},
  participates/.style={-{Stealth[length=6pt]}, thick, color=pathgreen},
  interacts/.style={-, color=geneblue!70, thick,
               bend left=18},
  attrlabel/.style={font=\scriptsize\color{slate}, align=left},
}

\title{Specification, Verification, and Synthesis over Heterogeneous Biomedical Knowledge Graphs}

\author{Jessica Louie}
\affiliation{%
  \institution{Spiralia LLC}
  \country{USA}
}
\email{jess@spiralia.co}
\author{Eric Palanques-Tost}
\affiliation{%
  \institution{Boston University}
  \country{USA}
}
\author{Ryan Matheu}
\affiliation{%
  \institution{University of Maryland}
  \country{USA}
}
\author{Kaan Certel}
\affiliation{%
  \institution{Spiralia LLC}
  \country{USA}
}
\author{Calin Belta}
\affiliation{%
  \institution{University of Maryland}
  \country{USA}
}
\email{calin@umd.edu}

\begin{document}

\begin{abstract}
We develop a formal framework for reasoning over heterogeneous Biomedical Knowledge Graphs (BKGs), motivated by cancer combination-target discovery. Existing formal methods for BKGs either query the graph or check its logical consistency; neither certifies a \emph{combinatorial selection} of BKG nodes against an optimization objective defined on external patient data, which is what target selection requires. We introduce Biomedical Knowledge-graph Modal Logic (\logicname{}), a two-layer modal logic: a \emph{graph layer} of typed, labeled, attributed node properties with edge and global modalities, and a \emph{selection layer} that lifts these to constraints on gene sets, including patient coverage. We define \emph{verification} (does a candidate gene set~$S$ satisfy a formula~$\varphi$?) and \emph{synthesis} (find a gene set, possibly after editing the BKG, that satisfies~$\varphi$ while maximizing patient coverage). We show that each formula compiles to a polynomial-size mixed-integer linear program (MILP): against a fixed BKG for verification and gene selection, and with size quadratic in the number of nodes when synthesis may edit the graph. We illustrate the framework using an oncology knowledge graph with 1{,}723 nodes and 4{,}502 edges and mutation data from 1{,}042 NSCLC patients, showing how progressively stronger biological specifications change the optimal target combination and achievable patient coverage.
\end{abstract}

\maketitle

\section{Introduction}
\label{sec:intro}

Combination therapies are increasingly important in cancer treatment, as targeting multiple molecular mechanisms can help overcome resistance to single-agent therapies. However, identifying promising combinations requires determining not only which genes are suitable therapeutic targets, but also which combinations are relevant to the patient population. The first question requires integrating biological evidence from multiple sources, including cancer-cell dependency, druggability, and molecular pathways. Biomedical knowledge graphs (BKGs) provide a structured way to represent and connect this heterogenuous evidence. The second question requires patient-level data to determine which patients carry alterations in the candidate genes. Although existing methods can address these questions separately, the challenge is to combine biological evidence from BKGs with patient data to select gene combinations that satisfy biological constraints while maximizing a population-level objective, such as patient mutation coverage.

This challenge is illustrated by MUC16, which is mutated in approximately 39\% of patients in our non-small cell lung cancer (NSCLC) cohort but is not classified as a cancer driver and shows little evidence of cancer-cell dependency. An optimization method that maximizes mutation coverage alone may favor MUC16 despite its limited justification as a therapeutic target. Conversely, requiring stronger biological evidence may exclude frequently mutated genes and reduce the achievable patient coverage. Selecting a combination therefore requires balancing biological requirements with population-level objectives. 

In this paper, we introduce Biomedical Knowledge-graph Modal Logic 
(\logicname{}), a formal language for specifying, verifying, and selecting gene combinations over heterogeneous biomedical knowledge graphs. The language allows users to express biological requirements involving gene attributes and relationships in the graph, together with constraints on the selected combination and patient coverage. We define Boolean and quantitative semantics for these requirements and provide an automatic translation into mixed-integer linear programs (MILPs). This enables the selection of gene combinations that satisfy the specified biological constraints while maximizing patient mutation coverage, with formal guarantees of correctness and optimality relative to the available data.

Our main contributions are:

\begin{enumerate}
\item \textbf{Specification language.}
We introduce \logicname{}, combining graph-based predicates with gene-set constraints and patient coverage, and establish robustness soundness and graph-layer decidability (Propositions~\ref{prop:soundness} and~\ref{prop:decidable}).

\item \textbf{Exact verification and synthesis.} We formalize both problems and derive a compositional, polynomial-size MILP encoding for the full language (Theorems~\ref{thm:milp} and~\ref{thm:milp-full}), with correctness and complexity guarantees.

\item \textbf{Biomedical applications.} We illustrate the framework using a public BKG, showing how graph-derived biological requirements can be combined with constraints on target selection.
\end{enumerate}

The remainder of the paper is organized as follows. 
Related work is reviewed in Section~\ref{sec:related}. 
Section~\ref{sec:kg} introduces the BKG as a formal object and the running NSCLC
example. Section~\ref{sec:spec} defines the syntax and semantics of
\logicname{}, its Kripke embedding, and discusses its expressivity.   Section~\ref{sec:verify} defines the
verification and synthesis problems and compiles them to MILPs, while Section~\ref{sec:encoding} gives the corresponding algorithms and discusses their complexity.  Section~\ref{sec:case-study} concludes our running example by solving target selection problems in NSCLC. The paper concludes with a discussion on open problems and future work in 
Section~\ref{sec:open}. 

\section{Related work}
\label{sec:related}

\paragraph{Knowledge graph querying and reasoning.} Languages such as SPARQL, Cypher, and regular path queries allow users to search knowledge graphs for nodes and connections satisfying specified conditions \cite{Barcelo13,CDLV00,Deutsch22}. Description logics provide additional tools for reasoning about the consistency and properties of knowledge bases~\cite{Baader03}. These approaches can identify genes with particular biological properties or relationships, but they do not directly address how to select an optimal combination of genes under constraints involving both the graph and patient data. Our framework connects these two tasks: graph-based conditions become part of the requirements used to select a gene combination. Integrative BKGs such as Hetionet~\cite{Himmelstein17}, PrimeKG~\cite{Chandak23}, and SPOKE~\cite{Morris23} prioritize drugs and targets with learned or path-based scores; they rank candidates but do not certify a selection against explicit constraints.

\paragraph{Combinatorial optimization.} Selecting combinations under constraints has been studied in databases and biomedical applications. Package queries~\cite{Brucato16} select groups of records that satisfy specified conditions while maximizing an objective. Similarly, Pang et al.~\cite{Pang14} use mixed-integer linear programming (MILP) to identify combinations of drug targets. These methods demonstrate that constrained selection can be formulated as an optimization problem. However, their constraints are not expressed through a language that directly supports navigation of heterogeneous BKGs. In contrast, \logicname{} allows biological requirements to be written as logical formulas involving node attributes and graph connections, which are then automatically translated into optimization constraints.

\paragraph{Logic-based specification and optimization.} Our language builds on modal logic, which describes properties of nodes and their connections in a graph \cite{Blackburn01,GorankoPassy92}. Previous work has also shown how logical requirements can be translated into MILP constraints~\cite{Raman14,Belta19}, and how numerical scores can measure whether these requirements are satisfied \cite{Donze10}. We build on these established ideas rather than introducing a new logic or optimization method. The main contribution is to combine them into a framework where biological evidence from a knowledge graph and requirements on gene combinations can be expressed, checked, and optimized together.

\section{Biomedical Knowledge Graph (BKG)}
\label{sec:kg}

\subsection{Definition}

We model the BKG as a typed, labeled, attributed multigraph. Types distinguish the roles of gene, pathway, and disease nodes;
real-valued attributes carry the evidence scores (mutation frequency, essentiality) that
determine target prioritization, which Boolean labels cannot; and labeled edges
carry the path-dependent properties (``is this gene associated with a disease
subtype through a driver \footnote{A \emph{driver} gene is one whose mutations help cause cancer, for example by letting cells divide without control. Tumors also carry many mutations that contribute nothing to their growth; genes hit only by such mutations are called \emph{passenger} genes. \label{fn:driver}} neighbor?'') that a flat table cannot.

Definition~\ref{def:kg} is parametric in a finite set $T$ of \emph{node types}, a
finite set $L$ of \emph{edge labels}, and the attribute domains $A_t$. For the
oncology graph, the paper's running example, we have 
\[
T = \{\textsc{Gene},\allowbreak\, \textsc{Disease},\allowbreak\, \textsc{Pathway}\}
\]
and
\begin{align*}
L = \{&\textsc{AssociatedWith},\, \textsc{ParticipatesIn},\, \textsc{ListedInKEGG},\\
&\textsc{InteractsWith},\, \textsc{SubtypeOf}\}.
\end{align*}

\begin{definition}[Biomedical Knowledge Graph]\label{def:kg}
A \emph{biomedical knowledge graph} is a tuple
$\KG = (V,\, E,\, \tau,\, \alpha)$ where:
\begin{itemize}[topsep=2pt,itemsep=1pt]
  \item $V$ is a finite set of \emph{nodes};
  \item $\tau : V \to T$ assigns each node a \emph{type};
  \item $E \subseteq V \times L \times V$ is a set of \emph{labeled directed edges}
        (a multigraph: multiple edges of different labels may connect the same pair of nodes);
  \item $\alpha : V \to \bigcup_{t \in T} A_t$ is a \emph{type-indexed attribute function},
        where $A_t$, $t\in T$ is an attribute domain, and $\alpha(v) \in A_{\tau(v)}$ for each $v \in V$. 
\end{itemize}
\end{definition}

\begin{remark}\label{rem:schema} Definition~\ref{def:kg} does not restrict which
node types a given edge label may connect. In the BKGs considered in this paper,
genes are \textsc{AssociatedWith} diseases. A gene \textsc{ParticipatesIn} a
pathway when it is a component of one of the ten oncogenic signaling pathways
curated for TCGA~\cite{SanchezVega18} (e.g., RTK--RAS, PI3K, TP53), and is
\textsc{ListedInKEGG} for a pathway when it is drawn on the KEGG pathway
map~\cite{Kanehisa23}; KEGG maps also include upstream regulators and downstream
effectors, so this edge is more permissive. Genes \textsc{InteractsWith} genes,
for the protein--protein interactions of STRING~\cite{Szklarczyk23}; these are
undirected, so $E$ contains both orientations. Finally, a disease can be a
\textsc{SubtypeOf} another (e.g., \textsc{LUAD} and \textsc{LUSC} of
\textsc{NSCLC}), which lets a query navigate from a subtype to its parent. No
other (type, label, type) triples occur.
\end{remark}

An \emph{attribute} is a named property recorded on every node of a given type,
such as a gene's mutation frequency or its essentiality score. Each node type~$t$
has a fixed list of attributes, and its \emph{attribute domain}~$A_t$ is the
Cartesian product of their value sets, so that $\alpha(v) \in A_{\tau(v)}$ is the
vector of $v$'s attribute values. In the displays below, the set above each brace
is an attribute's value set and the label under it is the attribute.
The attribute domain for each node type is: 
\[
A_{\textsc{Gene}} \;=\;
\underbrace{\R}_{\text{mut\_freq}} \times
\underbrace{\R}_{\text{essentiality}} \times
\underbrace{\bools}_{\text{driver}} \times
\underbrace{\bools}_{\text{druggable}} \times
\underbrace{\bools}_{\text{cancer\_pathway}}
\]
\[
A_{\textsc{Pathway}} \;=\; \Sigma^* \times \bools
\quad(\text{pathway id},\; \text{cancer\_relevant})
\]
\[
A_{\textsc{Disease}} \;=\; \Sigma^* \times \Sigma^*
\quad(\text{DOID code},\; \text{name})
\]
where $\Sigma$ is a set of alpha numeric symbols and $\Sigma^*$ is the set of all words over $\Sigma$. 

The \textsc{Gene} attributes are listed below. The label in parentheses at the end of an item, C2 through C4, names the constraint of the coverage optimization problem (see Example~\ref{ex:milpspec}) that reads that attribute; the full constraint set C1--C6 is defined in
Table~\ref{tab:c1c8} (Section~\ref{sec:twolayer}).
\begin{itemize}[topsep=2pt,itemsep=1pt]
  \item \texttt{mut\_freq}: the fraction of patients in the disease cohort carrying a
        somatic mutation in the gene, from TCGA~\cite{TCGA14lung,TCGA12lusc,TCGA12breast};
  \item \texttt{essentiality}: the gene's CRISPR knockout effect score (Chronos)
        from DepMap~\cite{DepMap24}; more negative means cancer cells depend more
        on the gene, and $\text{essentiality} < -0.5$ counts as essential (C2);
  \item \texttt{driver}: $1$ if the gene is a cancer driver according to the
        IntOGen compendium~\cite{Martinez20}, $0$ otherwise;
  \item \texttt{druggable}: $1$ if a drug is known to act on the gene's product,
        from DGIdb~\cite{Cannon24} (C3); 
  \item \texttt{cancer\_pathway}: $1$ if the gene participates in one of the
        curated oncogenic signaling pathways, that is, has a
        \textsc{ParticipatesIn} edge (C4).
\end{itemize}
A \textsc{Pathway} node records its identifier (a string: a KEGG identifier
such as \texttt{hsa04012}, or the name of a curated pathway such as
\texttt{RTK-RAS}) and a flag that is $1$ if the pathway is cancer-relevant:
always for the curated pathways, and for a KEGG pathway unless KEGG classifies
it under a non-cancer disease.
A \textsc{Disease} node records its Disease Ontology identifier (DOID) and its name.

For a node $v$ and attribute $a$ of its type, we write $\alpha(v, a)$ for the value of $a$ at~$v$, i.e., the $a$-component of the vector $\alpha(v)$.
Strictly, $\alpha$ takes one argument, so $\alpha(v,a)$ is shorthand for
$\pi_a(\alpha(v))$, where $\pi_a$ projects a vector in $A_{\tau(v)}$ onto its
$a$-component; we use the two-argument form throughout for readability.
We write $\VGene = \{v \in V : \tau(v) = \textsc{Gene}\}$ for the set of genes in the graph,
and similarly $V_{\textsc{Pathway}},\, V_{\textsc{Disease}}$.

\begin{remark}[Numeric and string attributes]
\label{rem:attrnames-split}
Attributes are either numeric (real or Boolean, such as essentiality and
driver status) or string-valued (such as pathway identifiers and DOID codes):
$\mathrm{Attr} = \mathrm{NumAttr} \cup \mathrm{StrAttr}$. Order comparisons
$\alpha(v,a)\,\theta\,c$ are well-formed only for $a \in \mathrm{NumAttr}$;
string-valued attributes get only an equality atomic formula. Throughout, $v$ in
an atomic formula denotes the node at which the formula is evaluated, not a
bound variable; inside $\exists g\in S.\,\varphi_G(g)$ that node is~$g$.
\end{remark}

Our current BKG has $|V| = 1{,}723$ (1{,}470 genes, 237 pathways, 16 diseases),
$|E| = 4{,}502$ (2{,}843 \textsc{AssociatedWith}, 104 \textsc{ParticipatesIn},
957 \textsc{ListedInKEGG}, 583 \textsc{InteractsWith} interactions,
15 \textsc{SubtypeOf}), with 127 genes
shared across NSCLC and HER2$^+$ breast cancer. It spans ten cancer types; genes shared across three or
more of the ten cancers number at 281.
This count uses genes annotated as drivers rather than the most frequently
mutated genes.  Including most frequently mutated passenger (see footnote~\ref{fn:driver}) genes that are not relevant to tumorigenesis,  such as
\textsc{MUC16}, in this shared category inflates the apparent overlap between cancer types.
This is an instance of the passenger-gene problem that motivates
Example~\ref{ex:nsclc} below. 

\begin{figure*}[t]
\centering
\begin{tikzpicture}

  \node[disease] (nsclc) at (0,3.2) {NSCLC};

  \node[gene]      (egfr) at (-6,0)  {EGFR\\{\tiny d=1,dg=1}};
  \node[gene]      (kras) at (-3,0)  {KRAS\\{\tiny d=1,dg=1}};
  \node[gene]      (pik)  at (0,0)   {PIK3CA\\{\tiny d=1,dg=1}};
  \node[gene]      (tp53) at (3,0)   {TP53\\{\tiny d=1,dg=1}};
  \node[passenger] (muc)  at (6.3,0) {MUC16\\{\tiny d=0,dg=0}};

  \node[pathway] (rtkras) at (-4.5,-4.2) {RTK--RAS\\signaling};
  \node[pathway] (pi3k)   at (0,-4.2)    {PI3K\\signaling};
  \node[pathway] (tp53p)  at (3,-4.2)    {TP53\\signaling};

  \draw[assocwith] (egfr) -- (nsclc);
  \draw[assocwith] (kras) -- (nsclc);
  \draw[assocwith] (pik)  -- (nsclc);
  \draw[assocwith] (tp53) -- (nsclc);
  \draw[assocwith, color=passgray] (muc) -- (nsclc);

  \draw[participates] (egfr) -- (rtkras);
  \draw[participates] (kras) -- (rtkras);
  \draw[participates] (pik)  -- (pi3k);
  \draw[participates] (tp53) -- (tp53p);

  \draw[interacts] (egfr) -- (kras);
  \draw[interacts] (kras) -- (pik);
  \draw[interacts] (pik)  -- (tp53);
  \draw[interacts] (egfr) to[bend right=40] (pik);
  \draw[interacts] (kras) to[bend right=40] (tp53);
  \draw[interacts] (egfr) to[bend right=46] (tp53);

  \begin{scope}[shift={(0,-6.6)}]
    \node[font=\small\bfseries\color{navy}] at (0,0.6) {Legend};
    \draw[assocwith]    (-7.0,0) -- (-6.2,0)
      node[right,font=\small\color{slate}]{\textsc{AssociatedWith}};
    \draw[participates] (-2.7,0) -- (-1.9,0)
      node[right,font=\small\color{slate}]{\textsc{ParticipatesIn}};
    \draw[interacts] (1.4,0) -- (2.2,0)
      node[right,font=\small\color{slate}]{\textsc{InteractsWith}};
    \node[gene, minimum size=0.5cm, text width=0pt, inner sep=0pt, font=\tiny, at={(-7.0,-0.9)}] {};
    \node[font=\small\color{slate}, anchor=west] at (-6.7,-0.9) {Driver gene};
    \node[passenger, minimum size=0.5cm, text width=0pt, inner sep=0pt, font=\tiny, at={(-4.2,-0.9)}] {};
    \node[font=\small\color{slate}, anchor=west] at (-3.9,-0.9) {Passenger gene};
    \node[disease, minimum size=0.5cm, font=\tiny, at={(-0.9,-0.9)}, aspect=1.8] {};
    \node[font=\small\color{slate}, anchor=west] at (-0.6,-0.9) {Disease};
    \node[pathway, minimum size=0.4cm, minimum width=0.9cm, font=\tiny, at={(1.4,-0.9)}] {};
    \node[font=\small\color{slate}, anchor=west] at (1.9,-0.9) {Pathway};
  \end{scope}

\end{tikzpicture}
\caption{%
  A subgraph of the NSCLC knowledge graph: five genes, three of the curated
  oncogenic signaling pathways~\cite{SanchezVega18}, and the NSCLC disease
  node, with every edge of the graph among these nodes. Gene
  nodes are annotated with \emph{d} (driver) and \emph{dg} (druggable); values
  are those of Table~\ref{tab:genes}. MUC16 (gray) is a passenger gene
  (driver~$=0$) despite being mutated in 39\% of the cohort. Edge color and
  style indicate the label (\textsc{AssociatedWith}: dashed red;
  \textsc{ParticipatesIn}: solid green; \textsc{InteractsWith}: solid blue,
  undirected). The full graph has 1{,}723 nodes and 4{,}502 edges
  (Remark~\ref{rem:schema}).
}
\label{fig:subgraph}
\end{figure*}
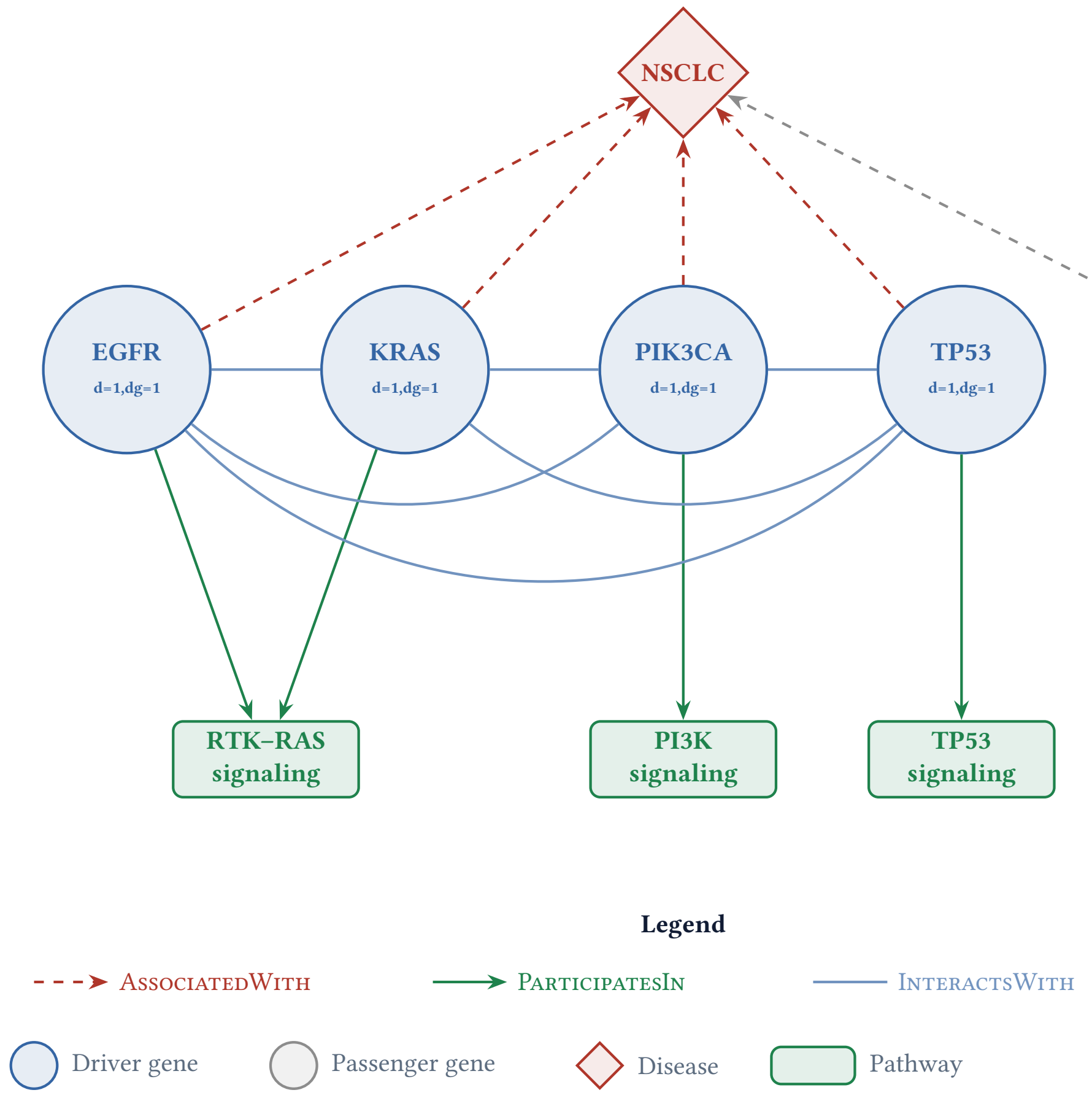

\begin{example}[NSCLC driver subgraph]\label{ex:nsclc}
Figure~\ref{fig:subgraph} depicts a subgraph of our BKG
restricted to five genes, three pathways, and the NSCLC (non-small cell lung
cancer) disease node. The heterogeneity of attributes is central: numeric evidence scores
(mutation frequency, essentiality),
Boolean selectors (driver status, druggability),
and ontology-typed identifiers (DOID codes, pathway identifiers)
coexist on the same graph. Table~\ref{tab:genes} lists the five genes' attributes for the NSCLC disease node, all five components of $A_{\textsc{Gene}}$, computed from the current graph and the TCGA NSCLC cohort of 1{,}042 patients.

\begin{table}[t]
\centering
\small
\setlength{\tabcolsep}{4pt}
\begin{tabular}{@{}l r r c c c@{}}
\toprule
Gene & mut\_freq & essentiality & driver & druggable & cancer\_pathway \\
\midrule
EGFR    & 0.081 & $-0.36$ & 1 & 1 & 1 \\
KRAS    & 0.151 & $-0.79$ & 1 & 1 & 1 \\
TP53    & 0.646 & $+0.25$ & 1 & 1 & 1 \\
PIK3CA  & 0.076 & $-0.48$ & 1 & 1 & 1 \\
MUC16   & 0.393 & $-0.05$ & 0 & 0 & 0 \\
\bottomrule
\end{tabular}
\caption{Genes in the NSCLC graph and their attributes.}
\label{tab:genes}
\end{table}

\noindent
Note that MUC16 is mutated in 39\% of the cohort, a high background rate
attributable to gene length rather than oncogenic selection: it is not a driver,
and cancer cells do not depend on it (essentiality near~$0$).
This disagreement between mutation frequency and driver evidence is
the ``passenger gene problem'' that motivates the specification language
in Section~\ref{sec:spec}: concretely, it is the disagreement between the BKG scorer
(which annotates MUC16 as a passenger and would exclude it) and an unconstrained
coverage optimizer (which would select high-frequency passengers such as MUC16
purely for their patient-coverage contribution if driver status is not encoded
as a constraint)~---~the ``MUC16
arbitration problem'' referenced throughout this paper.
\end{example}

\subsection{Patient data and the mutation matrix}

In addition to the BKG, we maintain a \emph{patient mutation matrix}.

\begin{definition}[Patient Mutation Matrix]\label{def:patient}
Given a set of $n$ candidate genes $U \subseteq \VGene$, numbered $1,\ldots,n$, and
a patient cohort $\mathcal{J} = \{1,\ldots,m\}$, the patient mutation matrix is
$P \in \bools^{n \times m}$, where $P_{ij} = 1$ iff patient~$j$ carries a
somatic mutation in gene~$i$.
\end{definition}

Our current dataset has $m_{\text{NSCLC}} = 1{,}042$ (557 LUAD, 485 LUSC) from TCGA whole-exome sequencing.

The pair $(\KG, P)$ is the complete input to the verification and synthesis
problems defined in Section~\ref{sec:problems}.


\section{Biomedical Knowledge-graph Modal Logic
(\logicname{})}
\label{sec:spec}

\medskip
The design principle of~\logicname{} is:
\emph{the graph layer specifies properties of individual nodes
(via types, attributes, and edge navigation),
and the selection layer specifies properties of combinations of nodes}.
This separates graph reasoning from combinatorial optimization.

\subsection{Syntax: the two-layer language}
\label{sec:twolayer}

\begin{definition}[Graph Layer $\varphi_G$]\label{def:graphlayer}
The \emph{graph layer} of \logicname{} consists of formulas generated by:
\begin{align}
\varphi_G \;\Coloneqq\; &
  \tau(v) = t \;\mid\;
  \alpha(v,a)\,\theta\, c \;\mid\;
  \alpha(v,a') = s \;\mid\;
  \langle \ell \rangle \,\varphi_G \;\mid\;
  [\ell] \,\varphi_G \;\mid\; \nonumber \\
  & \varphi^1_G \wedge \varphi^2_G \;\mid\;
  \neg\varphi_G \;\mid\;
  \mathsf{E}\,\varphi_G \;\mid\;
  \mathsf{A}\,\varphi_G \label{eq:graphlayer}
\end{align}
where $t \in T$, $a \in \mathrm{NumAttr}$, $\theta \in \{<,\le,=,\ge,>\}$,
$c \in \mathbb{Q}$, $a' \in \mathrm{StrAttr}$, $s \in \Sigma^*$, $\ell \in L$
(Remark~\ref{rem:attrnames-split}), and $\varphi^1_G$, $\varphi^2_G$ are formulas. 
$\neg$ and $\wedge$ are the Boolean operators negation and conjunction, respectively. 
$\langle\ell\rangle\varphi_G$ reads ``some $\ell$-successor satisfies
$\varphi_G$'' and $[\ell]\varphi_G$ ``every $\ell$-successor does'';
$\mathsf{E}\,\varphi_G$ reads ``some node of the graph satisfies $\varphi_G$''
and $\mathsf{A}\,\varphi_G$ ``every node does'' (the \emph{global
modalities}). 
\end{definition}

\begin{definition}[Selection Layer $\varphi_S$]\label{def:sellayer}
The \emph{selection layer} of \logicname{} lifts graph-layer properties to
constraints over a gene set $S \subseteq \VGene$:
\begin{align}
\varphi_S \;\Coloneqq\; &
  |S| \,\theta\, k \;\mid\;
  \exists g \in S.\,\varphi_G(g) \;\mid\;
  \forall g \in S.\,\varphi_G(g) \;\mid\; \nonumber \\
  & \varphi^1_S \wedge \varphi^2_S \;\mid\;
  \neg\varphi_S \;\mid\;
  \mathrm{cover}(S, P, \theta, c), \label{eq:sellayer}
\end{align}
where $k \in \mathbb{N}$, $\theta \in \{<,\le,=,\ge,>\}$, $c \in [0,1]$, and 
$\varphi^1_S$, $\varphi^2_S$ are formulas. We write $k_1 \le |S| \le k_2$ for $(|S|\ge k_1)\wedge(|S|\le k_2)$. In both (\ref{eq:graphlayer}) and (\ref{eq:sellayer}), the other Boolean operators (e.g., disjunction $\vee$) can be defined using negation and conjunction in the usual way.

We write $\varphi^{\mathrm{flat}}$ for the \emph{flat fragment}:
formulas in~\eqref{eq:graphlayer} and \eqref{eq:sellayer} without modal operators
$\langle \ell \rangle, [\ell], \mathsf{E}, \mathsf{A}$. Let
\begin{equation}\label{eqn:coverage}
\mathrm{coverage}(S, P) \;\coloneqq\;
\frac{1}{m}\,\bigl|\{j \in \mathcal{J} : \exists i \in S,\; P_{ij} = 1\}\bigr| \;\in\; [0,1]
\end{equation}

be the \emph{fraction of the patient cohort covered by $S$} (patient~$j$ is
covered if some selected gene is mutated in them). The Boolean predicate is then
$\mathrm{cover}(S,P,\theta,c) \coloneqq \mathrm{coverage}(S,P) \,\theta\, c$.
\end{definition}
Throughout the rest of the paper, we call $\max_S \,\mathrm{coverage}(S,P)$ the {\em coverage optimization} problem.

\begin{table}[t]
\centering
\small
\renewcommand{\arraystretch}{1.2}
\begin{tabular}{@{}l p{0.8\columnwidth}@{}}
\toprule
\textbf{Constraint} & \textbf{Meaning} \\
\midrule
C1 & Combination size: $2 \le |S| \le 3$ \\
C2 & At least one selected gene is essential ($\text{essentiality} < -0.5$) \\
C3 & At least one selected gene is druggable \\
C4 & At least one selected gene participates in a cancer-relevant pathway \\
C5 & Patient~$j$ is covered only if a selected gene is mutated in patient~$j$ \\
C6 & All decision variables are binary \\
\bottomrule
\end{tabular}
\caption{A representative constraint set for coverage-based target selection,
referenced by Example~\ref{ex:milpspec} and Theorem~\ref{thm:milp}.}
\label{tab:c1c8}
\end{table}

\begin{example}[A coverage-based selection specification in \logicname{}]\label{ex:milpspec}
Table~\ref{tab:c1c8} lists a representative constraint set for coverage-based
target selection (C1--C6); the corresponding $\LG$ formula is:
\begin{align*}
\varphi_{\mathrm{cov}} \;\coloneqq\; &
  (2 \le |S| \le 3) \\
  & \wedge\; \exists g \in S.\;[\alpha(g,\text{essentiality}) < -0.5] \\
  & \wedge\; \exists g \in S.\;[\alpha(g,\text{druggable}) = 1] \\
  & \wedge\; \exists g \in S.\;[\alpha(g,\text{cancer\_pathway}) = 1] \\
  & \wedge\; \mathrm{cover}(S, P, \ge, 0)
\end{align*}
(the last conjunct is trivially true --- $\mathrm{coverage}(S,P) \ge 0$ always ---
and is included only so that $P$ is syntactically part of the constraint set). The optimization objective is
$\max_S \,\mathrm{coverage}(S,P)$: maximize the coverage fraction.
All five conjuncts are in $\varphi_S$ and reference only flat graph-layer
sub-formulas $\varphi_G^{\mathrm{flat}}$~---~they do not involve edge navigation.
Further requirements, such as safety constraints based on tissue expression,
are added the same way, as additional flat conjuncts. 
\end{example}

\begin{example}[The passenger gene constraint]
\label{ex:passenger}
The MUC16 arbitration problem (Example~\ref{ex:nsclc}) requires excluding
genes annotated as passengers.
A passenger gene predicate in \logicname{} is simply:
\[
  \varphi_{\mathrm{pass}}(g) \;\coloneqq\; [\alpha(g, \text{driver}) = 0]
  \;\wedge\; [\alpha(g, \text{essentiality}) \ge -0.5]
\]
(``$g$ is neither a driver nor a dependency'': the second conjunct is the negation
of C2's essentiality condition).
The extended constraint $\forall g \in S.\; \neg\varphi_{\mathrm{pass}}(g)$
added to $\varphi_{\mathrm{cov}}$ would exclude MUC16 from any feasible $S$.
This is a flat formula, so it compiles to MILP (Theorem~\ref{thm:milp}).

\emph{Current practice.} Driver annotations usually come from curated
compendia such as IntOGen~\cite{Martinez20}. These act as external labels: they
are only as complete as the compendium, and they carry no proof of
correctness.
\end{example}

\subsection{Semantics}
\label{sec:semantics}

Formulas of \logicname{} are interpreted directly over the pair $(\KG, P)$:
graph-layer formulas at a node of~$\KG$, selection-layer formulas at a
candidate gene set $S \subseteq \VGene$. We give both a Boolean and a
quantitative (robustness) semantics.

\begin{definition}[Boolean Semantics]\label{def:bool}
\emph{Graph layer.} The relation $\KG, v \models \varphi_G$, for a node $v \in V$,
is defined inductively on~$\varphi_G$:
\begin{align*}
\KG,v &\models \tau(v) = t
  &&\text{iff } \tau(v) = t \\
\KG,v &\models \alpha(v,a)\,\theta\, c
  &&\text{iff } a \text{ is an attribute of } \tau(v) \\
  &&&\quad\text{and } \alpha(v,a)\,\theta\, c \text{ holds} \\
\KG,v &\models \alpha(v,a') = s
  &&\text{iff } a' \text{ is an attribute of } \tau(v) \\
  &&&\quad\text{and } \alpha(v,a') = s \\
\KG,v &\models \langle \ell \rangle\,\varphi_G
  &&\text{iff } \KG,u \models \varphi_G \text{ for some } u \\
  &&&\quad\text{with } (v,\ell,u) \in E \\
\KG,v &\models [\ell]\,\varphi_G
  &&\text{iff } \KG,u \models \varphi_G \text{ for every } u \\
  &&&\quad\text{with } (v,\ell,u) \in E \\
\KG,v &\models \mathsf{E}\,\varphi_G
  &&\text{iff } \KG,u \models \varphi_G \text{ for some } u \in V \\
\KG,v &\models \mathsf{A}\,\varphi_G
  &&\text{iff } \KG,u \models \varphi_G \text{ for every } u \in V
\end{align*}
with $\wedge$, $\vee$, $\neg$ interpreted as usual. An attribute atom is false
at a node whose type does not carry that attribute (e.g.\ a \textsc{Gene}
attribute read at a \textsc{Pathway} node reached by $\langle\ell\rangle$).
$\mathsf{E}\,\varphi_G$ and $\mathsf{A}\,\varphi_G$ have the same truth value at
every node. $[\ell]\,\varphi_G$ holds vacuously at a node with no outgoing
$\ell$-edge.

\emph{Selection layer.} The relation $(\KG, P, S) \models \varphi_S$, for a
gene set $S \subseteq \VGene$, is defined inductively on~$\varphi_S$:
\begin{align*}
(\KG,P,S) &\models (|S| \,\theta\, k)
  &&\text{iff } |S| \,\theta\, k \text{ holds in } \mathbb{N} \\
(\KG,P,S) &\models \exists g \in S.\,\varphi_G(g)
  &&\text{iff } \KG, g \models \varphi_G \text{ for some } g \in S \\
(\KG,P,S) &\models \forall g \in S.\,\varphi_G(g)
  &&\text{iff } \KG, g \models \varphi_G \text{ for every } g \in S \\
(\KG,P,S) &\models \mathrm{cover}(S,P,\theta,c)
  &&\text{iff } \mathrm{coverage}(S,P) \,\theta\, c
\end{align*}
with $\wedge$ and $\neg$ interpreted as usual.
\end{definition}

The robustness semantics adapts the signal temporal logic (STL) robustness
measure of Donz\'e and Maler~\cite{Donze10} from continuous signals over time to
attribute vectors over graph nodes, inheriting its min/max structure for
conjunction and disjunction and its anti-monotonicity under negation.

\begin{definition}[Robustness Semantics]\label{def:rob}
Robustness is defined inductively at the same two levels as
Definition~\ref{def:bool}: $\rob(\varphi_G, v, \KG, P) \in \R\cup\{\pm\infty\}$ for a
graph-layer formula at a node $v\in V$, and $\rob(\varphi_S, S, \KG, P)$ for a
selection-layer formula at a gene set $S\subseteq\VGene$. A numeric comparison
$x\,\theta\,c$, where $x$ is $\alpha(v,a)$ for $a\in\mathrm{NumAttr}$, or $|S|$,
or $\mathrm{coverage}(S,P)$, has robustness
\[
\rob(x\,\theta\,c) \;=\; \begin{cases}
    x - c & \theta \in \{\ge, >\} \\
    c - x & \theta \in \{\le, <\} \\
    -|x - c| & \theta = {=}
  \end{cases}
\]
For a Boolean attribute (e.g., \texttt{driver}, \texttt{druggable},
\texttt{cancer\_pathway}) compared with $=$,
for type atoms $\tau(v)=t$, for string atoms $\alpha(v,a')=s$, and for any
attribute atom at a node whose type lacks the attribute, there is no continuous
margin to measure, so robustness saturates to $+1$ if the atom holds
(Definition~\ref{def:bool}) and $-1$ otherwise. The connectives are included as follows:
\begin{align*}
\rob(\neg\varphi) &\;=\; -\rob(\varphi) \\
\rob(\varphi_1 \wedge \varphi_2) &\;=\; \min\bigl(\rob(\varphi_1),\rob(\varphi_2)\bigr) \\
\rob(\varphi_1 \vee \varphi_2) &\;=\; \max\bigl(\rob(\varphi_1),\rob(\varphi_2)\bigr) \\
\rob(\langle\ell\rangle\varphi_G, v) &\;=\; \max_{(v,\ell,u)\in E} \rob(\varphi_G, u) \\
\rob([\ell]\varphi_G, v) &\;=\; \min_{(v,\ell,u)\in E} \rob(\varphi_G, u) \\
\rob(\mathsf{E}\,\varphi_G, v) &\;=\; \max_{u\in V} \rob(\varphi_G, u) \\
\rob(\mathsf{A}\,\varphi_G, v) &\;=\; \min_{u\in V} \rob(\varphi_G, u) \\
\rob(\exists g \in S.\,\varphi_G(g), S) &\;=\; \max_{g \in S} \rob(\varphi_G, g) \\
\rob(\forall g \in S.\,\varphi_G(g), S) &\;=\; \min_{g \in S} \rob(\varphi_G, g)
\end{align*}
(suppressing the arguments $\KG, P$), with $\max\emptyset = -\infty$ and
$\min\emptyset = +\infty$, so that $\langle\ell\rangle$ with no successor is
violated and $[\ell]$ with no successor holds, matching Definition~\ref{def:bool}.
\end{definition}

\begin{proposition}[Soundness of robustness]\label{prop:soundness}
For every formula $\varphi$ of either layer and every point of evaluation
($v\in V$ for $\varphi_G$, $S\subseteq\VGene$ for $\varphi_S$):
if $\rob(\varphi) > 0$ then $\varphi$ holds, and if $\rob(\varphi) < 0$ then
$\varphi$ does not hold (Definition~\ref{def:bool}).
\end{proposition}

The proof, by simultaneous induction on $\varphi$, is in Appendix~\ref{app:props}.

A positive robustness score is therefore a quantitative certificate of
satisfaction and a negative one a certificate of violation. A score of
exactly~$0$ is inconclusive: $\alpha(v,a) \ge c$ with $\alpha(v,a) = c$ holds,
and $\alpha(v,a) > c$ with $\alpha(v,a) = c$ does not, yet both have
robustness~$0$. Saturated atoms put $\pm1$ on the same scale as the real-valued
margins of numeric attributes; when robustness enters an objective, the margins should be normalized to comparable ranges.

\subsection{Decidability via a Kripke embedding}
\label{sec:kripke}

Verification and synthesis (Section~\ref{sec:verify}) use
Definition~\ref{def:bool} directly; this subsection connects the graph layer to
standard modal logic. A knowledge graph $\KG$ is a multi-relational Kripke
structure $M_\KG$ (Appendix~\ref{app:kripke}, Definition~\ref{def:kripke}): its
states are the nodes, there is one accessibility relation $R_\ell$ per edge
label, and the atomic propositions true at a node are the type and attribute
atoms it satisfies. The graph layer is therefore the multi-modal logic
$\mathbf{K}_n$ ($n=|L|$)~\cite{Blackburn01} extended with the global modalities
$\mathsf{E},\mathsf{A}$~\cite{GorankoPassy92}. This identification settles
decidability, and the precompilation pass of Section~\ref{sec:encoding} is
ordinary bottom-up model checking on~$M_\KG$.

\begin{proposition}[Decidability of the graph layer]\label{prop:decidable}
Satisfiability of graph-layer formulas over the class of all knowledge graphs
(Definition~\ref{def:kg}), that is, whether some finite $\KG$ and node $v$
satisfy $\KG, v \models \varphi_G$, is decidable in exponential time.
\end{proposition}

Appendix~\ref{app:props} gives the proof: the graph layer reduces to
multi-modal logic $\mathbf{K}_n$ with global modalities, whose satisfiability is in
\textsc{ExpTime}.
Constraints on the schema, such as ``\textsc{ParticipatesIn} edges end at
\textsc{Pathway} nodes'', are themselves $\mathsf{A}$-formulas, so
satisfiability relative to a fixed schema reduces the same way.

\begin{remark}
\label{rem:explosion}
Evaluating a graph-layer formula at every node takes time linear in
$|\varphi_G|\cdot(|V|+|E|)$, since every modality is a single edge step and
there is no unbounded path operator. The hard part is choosing the set: at
genome-wide scale ($|\VGene| \approx 19{,}000$) there are
$\binom{19{,}000}{3} \approx 1.1 \times 10^{12}$ sets of three genes, which is why
the MILP encoding of Section~\ref{sec:verify} is needed
(Remark~\ref{rem:data-complexity} gives the full complexity picture).
\end{remark}

\subsection{Biologically meaningful queries: flat vs.\ full language}
\label{sec:more-examples}

The examples so far (Examples~\ref{ex:milpspec}--\ref{ex:passenger}) are flat
and use the NSCLC subgraph of Figure~\ref{fig:subgraph}. We add two more, from
drug classes currently active in oncology. Example~\ref{ex:kras-flat} (KRAS(ON)
inhibition) is flat and uses the same NSCLC graph, with the values of
Table~\ref{tab:genes}. Example~\ref{ex:wrn-full} (WRN inhibition, a
synthetic-lethal dependency in a subtype of colorectal cancer identified by a
biomarker) is chosen because it cannot be expressed in the flat fragment at all.
It needs nodes the graph does not contain (\textsc{WRN} and an MSI-high disease
subtype), so it is stated over a schema-compatible extension of the graph and is
not evaluated. 

\begin{example}[Flat: KRAS druggability, and why an  attribute can go stale]
\label{ex:kras-flat}
In the subgraph shown in Figure \ref{fig:subgraph},  $\alpha(\textsc{KRAS},\text{druggable})=1$
(Table~\ref{tab:genes}). However, through most of the 2010s KRAS was the textbook ``undruggable'' oncogene and the correct value was~$0$. G12C
inhibitors reached the clinic in 2021, and as of 2026 Revolution Medicines'
zoldonrasib (RMC-9805), a covalent KRAS G12D-selective molecular glue, has
reported 52\% ORR in previously-treated KRAS G12D NSCLC, with the
broader pan-RAS(ON) daraxonrasib (RMC-6236) now FDA-NDA-accepted after a
Phase~3 plenary win. A flat query
\begin{align*}
\varphi_{\mathrm{KRAS}} \;\coloneqq\; \exists g\in S.\bigl[&\alpha(g,\text{driver})=1 \\
&\;\wedge\; \alpha(g,\text{essentiality})<-0.5 \\
&\;\wedge\; \alpha(g,\text{druggable})=1\bigr]
\end{align*}
is well-formed and flat, and holds at $S=\{\textsc{KRAS}\}$ on the current graph
($\text{essentiality}=-0.79$); on a graph built before 2021 it would not, because
the \texttt{druggable} bit would be~$0$. This is not a defect in the flat
fragment or in $\varphi_{\mathrm{KRAS}}$~---~it is a reminder that
Theorem~\ref{thm:milp}'s guarantee is only as good as $\alpha$: the
compilation is sound relative to \emph{whatever} $\KG$ it is given, and
keeping druggability annotations current as new modalities reach the clinic
is a KG-maintenance problem the language itself cannot solve.
\end{example}

\begin{example}[Full language, not flat: WRN essentiality is conditional on a biomarker the gene itself does not carry]
\label{ex:wrn-full}
WRN helicase inhibition is a validated synthetic-lethal dependency, but
specifically and only in mismatch-repair-deficient (MSI-high) tumors; in
microsatellite-stable tumors WRN is not essential. This is exactly the kind
of dependency a single flat $\alpha(\textsc{WRN},\text{essentiality})$ number cannot
represent: essentiality here is not a property of the gene alone, it
is a property of the (gene, tumor-subtype) pair, and tumor subtype lives on
a \emph{different} node (a \textsc{Disease} node, reached only by edge
navigation), not on \textsc{WRN}'s own attribute vector. Using the existing
\textsc{SubtypeOf} hierarchy (Remark~\ref{rem:schema} already establishes
\textsc{LUAD}/\textsc{LUSC} as subtypes of \textsc{NSCLC}; MSI-high
colorectal cancer is, in exactly the same way, a recognized molecular
subtype of colorectal cancer, so this reuses existing schema, no new node
type required):
\begin{align*}
&\varphi_{\mathrm{WRN}}(g) \;\coloneqq\; \langle\textsc{AssociatedWith}\rangle\Bigl(\\
  &\quad\alpha(v,\text{name})=\text{``MSI-high colorectal cancer''} \\
  &\quad\wedge\; \langle\textsc{SubtypeOf}\rangle\bigl(\alpha(v,\text{name})=\text{``colorectal cancer''}\bigr)\Bigr)
\end{align*}
(``$g$ is associated with a disease named `MSI-high colorectal cancer' that is
a subtype of colorectal cancer''; \textsc{SubtypeOf} edges point from subtype
to parent, as in Remark~\ref{rem:schema}).
This has modal nesting depth~2 in Theorem~\ref{thm:milp-full}'s
sense, and is not expressible in $\varphi_G^{\mathrm{flat}}$
(Definition~\ref{def:graphlayer}) at all, by construction: the flat fragment
has no modal operators, full stop, so there is no
flat rewriting of ``depends on which disease subtype you navigate to,''
regardless of how the attribute vocabulary is extended. It \emph{is}
MILP-compilable, per Theorem~\ref{thm:milp-full}: at modal depth~2, on the
graph extended with the \textsc{WRN} and MSI-high nodes, it precompiles in a
bottom-up pass before the MILP is built (Appendix~\ref{app:milpproof-full}), no
different in kind from Example~\ref{ex:kras-flat}, just requiring navigation
Example~\ref{ex:kras-flat} does not.
\end{example}

\begin{remark}[Queries that need the full language]
\label{rem:flat-vs-full}
The flat fragment evaluates a gene's own attributes but cannot
directly navigate graph relationships. Two examples requiring the
full language are (i) association with a specific disease subtype
(Example~\ref{ex:wrn-full}) and (ii) participation in PI3K signaling
(Section~\ref{sec:case-study}). Both require edge modalities and
compile to MILP by Theorem~\ref{thm:milp-full}. On a fixed BKG,
these graph predicates can be precomputed without increasing
the size of the resulting MILP.

However, some requirements remain inexpressible in \logicname{}.
For example, the language cannot generally specify that two
selected genes participate in the same pathway, since it cannot
compare the identities of nodes reached from different genes
(Section~\ref{sec:open}).
\end{remark}

\section{Verification and Synthesis}
\label{sec:verify}

The verification and synthesis problems defined here use the 
BKG $\KG$ and the patient matrix $P$ defined in Section \ref{sec:kg}. 

\subsection{The verification and synthesis problems}
\label{sec:problems}

\begin{problem}[Verification]\label{prob:verify}
Given $\KG$, $P$, a BKML formula $\varphi_S$, and a candidate gene set
$S \subseteq \VGene$, (a) (Boolean) decide whether $(\KG, P, S) \models \varphi_S$
(Definition~\ref{def:bool}); (b) (Quantitative) Compute
$\rob(\varphi_S, S, \KG, P)$ (Definition~\ref{def:rob}).
\end{problem}

To define the synthesis problem, we need to define allowed operations on a BKG $\KG$. In the biomedical context, operations correspond to genetic or pharmacological interventions. We assume that synthesis may also edit the graph, with five operations defined in
Appendix~\ref{app:ops} (Definition~\ref{def:ops}): \textbf{Delete}$(v)$ removes a
node and its edges (gene knockout); \textbf{Attenuate}$(v,a,\delta)$ and
\textbf{Agonize}$(v,a,\delta)$ lower or raise a numeric attribute by
$0 \le \delta \le \delta_{\max}$ (partial suppression or activation); and
\textbf{AddEdge} and \textbf{RemoveEdge} insert or delete a labeled edge (a newly
found interaction, or a reclassified association). An operation sequence
$\mathcal{O} = (o_1, \ldots, o_k)$ yields $\mathcal{O}(\KG)=\KG' = o_k \circ \cdots \circ o_1(\KG)$,
with $|\mathcal{O}| = k$.

\begin{problem}[Synthesis]\label{prob:synth}
Given $\KG$, a BKML formula $\varphi_S$, a patient matrix $P$, a budget
$K \in \mathbb{N}$, 
and a finite set of permitted operations $\mathcal{O}$, find a gene set
$S^*$ and a sequence $\mathcal{O}^*$ of permitted operations with
$|\mathcal{O}^*| \le K$ such that
\[
(\KG', P, S^*) \models \varphi_S
\]
and $(S^*, \mathcal{O}^*)$
maximizes $\mathrm{coverage}(S^*, P)$ (Equation~\eqref{eqn:coverage}) over all $(S, \mathcal{O})$. In the above, $\KG' = \mathcal{O}^*(\KG)$.
\end{problem}

With $K = 0$ (no BKG modifications permitted), $\KG' = \KG$ and Problem~\ref{prob:synth} is exactly the coverage optimization problem.

A solution to either problem is certified relative to its specification: the
selected set provably satisfies $\varphi_S$ over $(\KG, P)$ and, for synthesis,
maximizes coverage among the sets that do. Whether the criteria encoded in
$\varphi_S$ predict therapeutic benefit is a scientific and clinical question
outside the logic, to be settled experimentally and ultimately in clinical
trials.

\subsection{Compilation to MILP}

The central algorithmic result is that every BKML formula compiles to a
polynomial-size MILP, which serves both verification (checking a fixed~$S$)
and synthesis (optimizing over~$S$). We state the flat fragment first, since
it needs no graph traversal. Throughout, $q$ is the number of
$\exists g\in S$ and $\forall g\in S$ quantifiers in $\varphi_S$, and
$\mathrm{nnz}(P)$ is the number of entries of $P$ equal to~$1$.

\begin{theorem}[MILP Compilation]\label{thm:milp}
Let $\varphi_S$ have all graph-layer sub-formulas in
$\varphi_G^{\mathrm{flat}}$. There is a MILP $\Pi(\varphi_S, \KG, P)$ over binary
variables $x_i$ ($i\in\VGene$), $z_j$ ($j\in\mathcal{J}$), and one $b_\psi$ per
selection-layer sub-formula $\psi$ of $\varphi_S$, so
$O(|\VGene| + m + |\varphi_S|)$ binary variables, with
$O(q\cdot|\VGene| + \mathrm{nnz}(P) + |\varphi_S|)$ linear constraints, such that
for every $S \subseteq \VGene$:
\begin{multline*}
\Pi \text{ with } x \text{ fixed to the indicator vector of } S
\text{ is feasible} \\ \;\Longleftrightarrow\; (\KG, P, S) \models \varphi_S ,
\end{multline*}
and in every feasible solution $z_j = 1$ exactly when patient~$j$ is covered by~$S$.
Hence $\Pi$ is feasible if and only if some $S$ satisfies $\varphi_S$, and maximizing the
linear objective $\frac1m\sum_j z_j$ over $\Pi$ solves the zero-operation case
of Problem~\ref{prob:synth}. The compilation runs in time
$O(|\varphi_S|\cdot|\VGene| + \mathrm{nnz}(P))$.
\end{theorem}

\begin{proof}[Proof sketch]
Attribute values are fixed data, so each flat graph-layer sub-formula has a
fixed truth value $c_i$ at each gene~$i$. Each selection-layer sub-formula
then gets an exact binary indicator from standard gates (OR over the selected
genes with $c_i=1$ for $\exists g\in S$, its dual for $\forall g\in S$, AND for
$\wedge$, $1-b$ for $\neg$, and two-sided bounds for size and coverage), and
induction on the formula shows that each indicator equals the truth value.
The full construction is provided in  Appendix~\ref{app:milpproof} and the algorithm is given in
Section~\ref{sec:encoding}.
\end{proof}

\begin{remark}[How much of the language of interest is already flat]
\label{rem:flat-coverage}
Every constraint of Table~\ref{tab:c1c8} (C1--C4 are the conjuncts of
$\varphi_{\mathrm{cov}}$; C5 and C6 belong to the encoding) and the passenger
predicate $\varphi_{\mathrm{pass}}$ are flat, so Theorem~\ref{thm:milp} already
covers the whole specification of the coverage optimization problem. Non-flat predicates, such
as the pathway-adjacency check
$\langle\textsc{ParticipatesIn}\rangle(\tau(v)=\textsc{Pathway})$, are covered by
Theorem~\ref{thm:milp-full} below.
\end{remark}

\begin{theorem}[Full-Language MILP Compilation]\label{thm:milp-full}
Let $\varphi_S$ be any formula of the full BKML
(Definitions~\ref{def:graphlayer}--\ref{def:sellayer}). 
\begin{enumerate}[topsep=2pt,itemsep=2pt,label=(\roman*)]
\item \textbf{Fixed BKG} (verification, and Problem~\ref{prob:synth}
with $K=0$): after a precompilation pass taking $O(|\varphi_S|\cdot(|V|+|E|))$
time, there is a MILP $\Pi(\varphi_S,\KG,P)$ with exactly the size and
correctness guarantees of Theorem~\ref{thm:milp}.
\item \textbf{Symbolic BKG} (Problem~\ref{prob:synth} with $K>0$):
there is a MILP, with one continuous variable per permitted Attenuate/Agonize
pair, $O(|\varphi_S|\cdot|V|^2 + |\VGene| + m + N_{\mathrm{op}})$ variables and
$O(|\varphi_S|\cdot|V|^2 + \mathrm{nnz}(P) + N_{\mathrm{op}})$ constraints, where
$N_{\mathrm{op}}$ is the number of permitted operations, whose feasible
solutions correspond exactly to the feasible pairs $(S,\mathcal{O})$ of
Problem~\ref{prob:synth}. Strict comparisons on attributes that an operation can
change are read with a tolerance $\varepsilon > 0$ ($x > c$ as
$x \ge c + \varepsilon$)~\footnote{this is a standard assumption for MILP encodings.}. 
\end{enumerate}
In both cases the MILP has polynomial size; in case~(ii) the size no longer
depends on modal nesting depth.
\end{theorem}

\begin{proof}[Proof sketch]
\emph{Case (i).} On a fixed graph, a bottom-up pass evaluates every
graph-layer sub-formula at every node, in $O(|V|+|E|)$ per sub-formula; this
yields constants $c_i$ at the genes, and the construction of
Theorem~\ref{thm:milp} applies unchanged.
\emph{Case (ii).} An edit can change truth values, so the truth value of each
sub-formula at each node becomes a binary variable, tied to its value in the
edited graph by exact gates; a modal sub-formula ranges over at most $|V|^2$
successor pairs, which gives the quadratic term. The full construction can be found in Appendix~\ref{app:milpproof-full}.
\end{proof}

\begin{remark}[Complexity]
\label{rem:data-complexity}
\emph{Verification} (Problem~\ref{prob:verify}) is in polynomial time: evaluate
Definition~\ref{def:bool} directly, in $O(|\varphi_S|\cdot(|V|+|E|) + |S|\cdot m)$.
\emph{Synthesis} (Problem~\ref{prob:synth}) is NP-hard, already for
$\varphi_S = (|S|\le k)$ with $K=0$, since that is maximum
coverage~\cite{Hochbaum97}.. 
\end{remark}

\section{The Encoding Scheme}
\label{sec:encoding}

Theorems~\ref{thm:milp} and~\ref{thm:milp-full} establish \emph{that} BKML
compiles to MILP; this section gives the compilation for a fixed knowledge
graph as two algorithms on the formula syntax tree, given in full in
Appendix~\ref{app:algorithms}. Algorithm~\ref{alg:precompile}
evaluates the graph layer; Algorithm~\ref{alg:full} builds the MILP from the
selection layer. The Python encoder accompanying this paper implements both.
Throughout, $|\psi|$ denotes the number of syntax-tree nodes of a
(sub)formula~$\psi$.

\subsection{Precompiling the graph layer}

On a fixed graph, a graph-layer formula has a fixed truth value at every node,
so it contributes no decision variables. Algorithm~\ref{alg:precompile} computes
these values bottom-up over the sub-formulas, filling a table
$\mathit{val}[\psi][v]$ for every sub-formula $\psi$ and node $v$. The global
modalities are computed once, since their value does not depend on the node.
For a flat formula, only the rows for genes are needed and no edge is read, so
the cost drops to $O(|\varphi_G|\cdot|\VGene|)$.
Each sub-formula costs $O(|V|)$, except $\langle\ell\rangle$ and $[\ell]$, which
read each $\ell$-edge once, so the total is $O(|\varphi_G|\cdot(|V|+|E|))$. 

\subsection{Compiling the selection layer}

Algorithm~\ref{alg:full} assigns every selection-layer sub-formula $\psi$ a
binary indicator $b_\psi$ constrained to equal $\psi$'s truth value for the
selected set, and then asserts $b_{\varphi_S}=1$. Since each indicator is
exact in both directions, negation is simply $1-b_\psi$, and no case of the
grammar is left out. Size and coverage comparisons use the integrality of
$\sum_i x_i \in \{0,\ldots,n\}$ and $\sum_j z_j \in \{0,\ldots,m\}$; every
comparator reduces to the two shown, since
$y\,{>}\,k \equiv y \ge k+1$, $y\,{\le}\,k \equiv \neg(y \ge k+1)$,
$y\,{<}\,k \equiv \neg(y\ge k)$, and $y = k$ is the conjunction of $\ge$ and
$\le$. For coverage, $\mathrm{coverage}(S,P)\ge c$ is $\sum_j z_j \ge \lceil cm\rceil$
and $\mathrm{coverage}(S,P) > c$ is $\sum_j z_j \ge \lfloor cm\rfloor + 1$.

For a size or coverage atom with $k\ge1$ (resp.\ $t \ge 1$), $b=1$ forces the
sum to at least the threshold and $b=0$ forces it below; a threshold of~$0$ is
always true, and the constraints then force $b=1$. When every $\mathrm{cover}$
atom occurs positively with $\ge$ or $>$, as in $\varphi_{\mathrm{cov}}$, the
lower bounds $z_j \ge x_i$ can be dropped: an optimal or feasible solution can
always raise $z_j$ to its true value. They are needed only for negated
$\mathrm{cover}$ or upper bounds on coverage.

\subsection{Complexity of the encodings}
\label{sec:complexity}

Table~\ref{tab:complexity} in Appendix~\ref{app:sizes} compares the encoding
sizes of the three cases, as stated in Theorems~\ref{thm:milp}
and~\ref{thm:milp-full}.

Note that, on a fixed graph, the flat and full languages produce MILPs of the same size:
all graph-layer structure, however deeply nested, is paid for once in
precompilation, outside the solver, and the MILP depends only on the
selection layer. A graph-navigating constraint is therefore no harder for the
solver than a flat one. Only structural synthesis (case~(ii)) brings the graph
itself into the MILP, at a cost quadratic in~$|V|$.

\subsection{Structural operations and scale}
\label{sec:synth}

Theorem~\ref{thm:milp-full}(ii) covers all five operations of
Definition~\ref{def:ops}: edge edits and deletions become binary variables, and
Attenuate and Agonize become bounded continuous variables~$\delta$, so the
solver chooses \emph{how much} to attenuate or agonize. The MILP is then
quadratic in~$|V|$ when modal sub-formulas read edited edges
(Table~\ref{tab:complexity}); restricting the permitted operations to the few
edges and attributes under consideration keeps this term small.

Coverage optimizers in practice run on curated sets of tens of genes.
At genome-wide scale ($|\VGene| \approx 19{,}000$), exact branch-and-bound may
become slow, and three standard strategies apply: LP relaxation and rounding,
repairing biological feasibility by projecting onto the constraint set; column
generation, adding genes to an active set by reduced cost; and pre-filtering
the genes by a prior ranking before running the MILP.

\section{Case Study: Target Selection in NSCLC}
\label{sec:case-study}

We illustrate BKML and the synthesis Problem \ref{prob:synth} with $K=0$ (i.e., the coverage optimization problem)~\footnote{Case studies using the full Synthesis Problem \ref{prob:synth}, where the optimization is performed over both gene sets and BKG operations, are left for future work.} using the NSCLC knowledge graph in Figure~\ref{fig:subgraph} and mutation data from $1042$  patients from the TCGA dataset~\cite{TCGA14lung,TCGA12lusc}. We consider five candidate genes (EGFR, KRAS, TP53, PIK3CA, and MUC16) and select combinations of two or three genes that maximize patient mutation coverage. We compare three specifications with progressively stronger biological requirements.

We first consider the specification $\varphi_{\mathrm{cov}}$ (Example~4.1), which requires at least one essential gene, one druggable gene, and one gene involved in a cancer-relevant pathway. The combination maximizing patient mutation coverage while satisfying $\varphi_{\mathrm{cov}}$ is ${\mathrm{KRAS},\mathrm{TP53},\mathrm{MUC16}}$, covering 844 patients ($81.0\%$). Note that MUC16 is not classified as a cancer driver, but is selected because it increases patient coverage by $72$ patients compared with selecting KRAS and TP53 alone.

We next exclude genes that satisfy the passenger predicate $\varphi_{\mathrm{pass}}$ (Example~4.2): $\varphi_1 = \varphi_{\mathrm{cov}} \wedge \forall g \in S.\,\neg\varphi_{\mathrm{pass}}(g)$. This excludes MUC16 and changes the optimal combination to $\{\mathrm{EGFR},\mathrm{KRAS},\mathrm{TP53}\}$, covering $804$ patients ($77.2\%$). Enforcing the additional biological requirement thus reduces mutation coverage by $3.8$ percentage points.

Finally, we require at least one selected gene to participate in PI3K signaling $\varphi_2 = \varphi_1 \wedge\exists g \in S.\,\langle\mathrm{ParticipatesIn}\rangle\bigl(\alpha(v,\mathrm{pathway\_id})=\mathrm{PI3K}\bigr).$  Unlike the previous constraints, this requirement explicitly uses graph relationships. We use the curated oncogenic pathway connections shown in Figure~\ref{fig:subgraph}, where PIK3CA is the only candidate connected to PI3K signaling. The optimal combination becomes $\{\mathrm{KRAS},\mathrm{TP53},\mathrm{PIK3CA}\}$, covering $790$ patients ($75.8\%$), a further reduction of 1.3 percentage points.

\begin{table}[t]
\centering
\caption{Optimal NSCLC target combinations under three
BKML specifications. Coverage is the fraction of patients
with a mutation in at least one selected gene.}
\label{tab:case-study}
\small
\begin{tabular}{@{}llr@{}}
\toprule
Specification & Optimal combination & Coverage \\
\midrule
Baseline (10) & KRAS, TP53, MUC16 & 81.0\% \\
Passenger exclusion (6) & EGFR, KRAS, TP53 & 77.2\% \\
Graph-based (3) & KRAS, TP53, PIK3CA & 75.8\% \\
\bottomrule
\end{tabular}
\end{table}

These results are summarized in Table~\ref{tab:case-study}, where the numbers in parentheses 
indicate the number of feasible combinations.
Adding biological constraints reduces this number
from 10 to 3 and produces a different optimal combination
at each step. Overall, mutation coverage decreases
from $81.0\%$ to $75.8\%$. This case study demonstrates
how BKML combines gene attributes, graph relationships,
and patient-level data to select combinations satisfying
explicit biological requirements while maximizing
patient mutation coverage.

\section{Conclusion and Future Work}
\label{sec:open}

We have presented \logicname{}, a two-layer modal logic for
heterogeneous biomedical knowledge graphs, and shown that it captures
the combination-target selection problem as a formal verification and
synthesis task. Every formula compiles to an exact MILP of polynomial size
(Theorems~\ref{thm:milp} and~\ref{thm:milp-full}): on a fixed knowledge graph
the MILP depends only on the selection layer, and structural synthesis
operations add a term quadratic in the number of nodes.
Section~\ref{sec:encoding} gives the compilation as algorithms with a matching
complexity analysis, and we provide a Python implementation of them. The
robustness semantics is sound (Proposition~\ref{prop:soundness}), and the Kripke
embedding places the graph layer in multi-modal logic with global modalities,
which makes its satisfiability decidable (Proposition~\ref{prop:decidable}).
Verification is polynomial, while synthesis is NP-hard and handled exactly by
the MILP (Remark~\ref{rem:data-complexity}).

Our future work can grouped into short-term and long-term goals. In short term, we will apply the method from this paper to an autoimmune knowledge graph. This will require some non-trivial extensions, including accommodating atomic formulas on edge attributes, and a selection objective that does not depend on a mutation matrix. We will also adapt our method to work for richer patient data. $P$ records only whether a gene carries a somatic mutation in a patient. Richer data exist for some patients (variant
allele frequency, copy number, expression, post-treatment profiles). 

In long term, we will focus on three main directions. First, we will extend the expressivity of BKML. For example, in its current form, BKML uses the global modalities $\mathsf{E},\mathsf{A}$ to let a graph-layer formula refer to
the whole graph while keeping it decidable (Proposition~\ref{prop:decidable}),
but they do not name nodes. Unlike the binders of hybrid logic~\cite{ABM99},
they cannot express that two modally reached nodes are the same, so ``no two
selected genes participate in a common pathway'' is out of reach. Adding a
binder would make it expressible at the price of decidability~\cite{ABM99}. Second, we will define temporal extensions for BKML. We will lift BKML from static (single-timepoint) to temporal specifications
that are checked over an observed or predicted treatment course, as
acquired resistance mutations emerge on-treatment (e.g., in about 15--20\% of NSCLC patients after osimertinib~\cite{Leonetti19}). Third, we will explore the genome-wide scalability of our approach. We will derive approximation ratios for the three strategies of
Section~\ref{sec:synth} at genome-wide scale ($|\VGene| \approx 19{,}000$).

\bigskip
\noindent
\textbf{Acknowledgments.}
The knowledge graph used in this paper was constructed from public data sources:
TCGA~\cite{TCGA14lung,TCGA12lusc,TCGA12breast},
KEGG~\cite{Kanehisa23},
the TCGA oncogenic signaling pathways~\cite{SanchezVega18},
STRING~\cite{Szklarczyk23},
DepMap~\cite{DepMap24},
DGIdb~\cite{Cannon24},
and IntOGen~\cite{Martinez20}.

The authors acknowledge the use of Anthropic Claude and OpenAI ChatGPT for assistance with manuscript preparation and software development. The authors have reviewed all generated material for technical correctness and accept full responsibility for the contents of this paper.


\clearpage
\appendix
\section{The compilation algorithms}
\label{app:algorithms}

Algorithms~\ref{alg:precompile} and~\ref{alg:full} are the two passes described in
Section~\ref{sec:encoding}.

\begin{algorithm}
\caption{\textsc{Precompile}$(\varphi_G, \KG)$}
\label{alg:precompile}
\begin{algorithmic}[1]
\Require graph-layer formula $\varphi_G$ (any, including $\langle\ell\rangle,[\ell],\mathsf{E},\mathsf{A}$); fixed $\KG=(V,E,\tau,\alpha)$
\Ensure $c : \VGene \to \bools$, the truth value of $\varphi_G$ at each candidate gene
\For{each sub-formula $\psi$ of $\varphi_G$, children before parents}
  \If{$\psi$ is an atom}
    \For{$v \in V$} $\mathit{val}[\psi][v] \gets [\KG, v \models \psi]$ \Comment{Definition~\ref{def:bool}} \EndFor
  \ElsIf{$\psi = \psi_1 \wedge \psi_2$, $\psi_1\vee\psi_2$, or $\neg\psi_1$}
    \For{$v \in V$} combine $\mathit{val}[\psi_1][v]$, $\mathit{val}[\psi_2][v]$ accordingly \EndFor
  \ElsIf{$\psi = \langle\ell\rangle\psi'$ or $[\ell]\psi'$}
    \For{$v \in V$} $\mathit{val}[\psi][v] \gets \bigvee$ (resp.\ $\bigwedge$) of $\mathit{val}[\psi'][u]$ over $(v,\ell,u)\in E$ \EndFor
  \ElsIf{$\psi = \mathsf{E}\,\psi'$ or $\mathsf{A}\,\psi'$}
    \State $b \gets \bigvee_{u\in V}$ (resp.\ $\bigwedge_{u\in V}$) $\mathit{val}[\psi'][u]$ \Comment{computed once}
    \For{$v \in V$} $\mathit{val}[\psi][v] \gets b$ \EndFor
  \EndIf
\EndFor
\State \Return $\{\, i \mapsto \mathit{val}[\varphi_G][i] : i \in \VGene \,\}$
\end{algorithmic}
\end{algorithm}

\begin{algorithm}
\caption{\textsc{Compile}$(\varphi_S, \KG, P)$}
\label{alg:full}
\begin{algorithmic}[1]
\Require $\varphi_S \in \LG$ (any); fixed $\KG$; patient matrix $P$
\Ensure MILP $\Pi$
\For{each distinct graph-layer sub-formula $\varphi_G$ of $\varphi_S$}
  \State $c[\varphi_G] \gets \textsc{Precompile}(\varphi_G, \KG)$ \Comment{Algorithm~\ref{alg:precompile}}
\EndFor
\State introduce $x_i \in \bools$ for $i \in \VGene$ and $z_j \in \bools$ for $j \in \mathcal{J}$
\State add $z_j \le \sum_i P_{ij} x_i$ and $z_j \ge x_i$ for every $j$ and every $i$ with $P_{ij}=1$ \Comment{$z_j$ = patient $j$ covered}
\State add $\textsc{Sel}(\varphi_S) = 1$; \Return $\Pi$
\Statex
\Function{Sel}{$\psi$} \Comment{returns an exact indicator of $\psi$}
  \If{$\psi = \neg\psi'$} \Return $1 - \textsc{Sel}(\psi')$ \EndIf
  \State introduce $b \in \bools$
  \If{$\psi = (|S| \ge k)$}
    \State add $\sum_i x_i \ge k\,b$ and $\sum_i x_i \le k - 1 + (n - k + 1)\,b$
  \ElsIf{$\psi = (\mathrm{coverage}(S,P) \ge c)$, threshold $t = \lceil cm \rceil$}
    \State add $\sum_j z_j \ge t\,b$ and $\sum_j z_j \le t - 1 + (m - t + 1)\,b$
  \ElsIf{$\psi = \exists g\in S.\,\varphi_G(g)$; let $C = \{i : c[\varphi_G](i) = 1\}$}
    \State add $b \ge x_i$ for $i \in C$, and $b \le \sum_{i\in C} x_i$
  \ElsIf{$\psi = \forall g\in S.\,\varphi_G(g)$; let $D = \{i : c[\varphi_G](i) = 0\}$}
    \State add $b \le 1 - x_i$ for $i \in D$, and $b \ge 1 - \sum_{i\in D} x_i$
  \ElsIf{$\psi = \psi_1 \wedge \psi_2$}
    \State $b_1 \gets \textsc{Sel}(\psi_1)$; $b_2 \gets \textsc{Sel}(\psi_2)$; add $b \le b_1$, $b \le b_2$, $b \ge b_1 + b_2 - 1$
  \EndIf
  \State \Return $b$
\EndFunction
\end{algorithmic}
\end{algorithm}

\section{The graph-editing operations}
\label{app:ops}

\begin{definition}[KG Operations]\label{def:ops}
The operation set $\Ops$ consists of:
\begin{description}[topsep=2pt,itemsep=2pt]
\item[\textbf{Delete}$(v)$.]
Remove node $v$ and all incident edges from $\KG$; a deleted gene can no
longer be selected, and $\mathsf{E},\mathsf{A}$ range only over remaining nodes.
\emph{Biological interpretation}: complete gene knockout
(CRISPR deletion, RNAi silencing).
\item[\textbf{Attenuate}$(v, a, \delta)$.]
Set $\alpha(v, a) \leftarrow \alpha(v,a) - \delta$ for numeric attribute~$a$,
with $0 \le \delta \le \delta_{\max}$ for a fixed bound $\delta_{\max}$.
\emph{Biological interpretation}: partial gene suppression
(shRNA knockdown reducing expression or activity).
\item[\textbf{Agonize}$(v, a, \delta)$.]
Set $\alpha(v, a) \leftarrow \alpha(v,a) + \delta$, with $0 \le \delta \le \delta_{\max}$.
\emph{Biological interpretation}: activation via small-molecule agonist
or expression upregulation.
\item[\textbf{AddEdge}$(u, \ell, v)$.]
Insert edge $(u, \ell, v)$ into $E$.
\emph{Biological interpretation}: newly discovered protein-protein interaction;
reclassification of a gene-disease association after new clinical evidence.
\item[\textbf{RemoveEdge}$(u, \ell, v)$.]
Delete edge $(u, \ell, v)$ from $E$.
\emph{Biological interpretation}: removal of a false-positive interaction;
reclassification of a passenger gene (removing its AssociatedWith edge to the disease node).
\end{description}
An \emph{operation sequence} is $\mathcal{O} = (o_1, \ldots, o_k)$ with
$o_i \in \Ops$; the resulting KG is
$\KG' = o_k \circ \cdots \circ o_1(\KG)$, and $|\mathcal{O}| = k$. None of the
operations adds a node.
\end{definition}

\section{Encoding sizes}
\label{app:sizes}

Let $q$ denote the number of $\exists g\in S$ and $\forall g\in S$ quantifiers in $\varphi_S$, and $\mathrm{nnz}(P)$ be the number of ones in~$P$.
Table~\ref{tab:complexity} (Appendix~\ref{app:sizes}) compares the encoding
sizes of the three cases, as stated in Theorems~\ref{thm:milp}
and~\ref{thm:milp-full}.

\begin{table*}[t]
\centering
\small
\renewcommand{\arraystretch}{1.3}
\begin{tabular}{@{}l l l l@{}}
\toprule
& \textbf{Flat} & \textbf{Full, fixed $\KG$} & \textbf{Full, symbolic $\KG$} \\
& (Thm.~\ref{thm:milp}) & (Thm.~\ref{thm:milp-full}.i) & (Thm.~\ref{thm:milp-full}.ii) \\
\midrule
Precompilation & $O(|\varphi_S|\cdot|\VGene|)$ & $O(|\varphi_S|\cdot(|V|+|E|))$ & --- \\
Binary variables & $O(|\VGene| + m + |\varphi_S|)$ & $O(|\VGene| + m + |\varphi_S|)$ & $O(|\varphi_S|\cdot|V|^2 + |\VGene| + m + N_{\mathrm{op}})$ \\
Constraints & $O(q\cdot|\VGene| + \mathrm{nnz}(P) + |\varphi_S|)$ & $O(q\cdot|\VGene| + \mathrm{nnz}(P) + |\varphi_S|)$ & $O(|\varphi_S|\cdot|V|^2 + \mathrm{nnz}(P) + N_{\mathrm{op}})$ \\
\bottomrule
\end{tabular}
\caption{Encoding size as a function of input size. $q\le|\varphi_S|$ counts the
$\exists/\forall$-over-$S$ quantifiers, $\mathrm{nnz}(P)$ the ones in~$P$, and
$N_{\mathrm{op}}$ the permitted operations of Problem~\ref{prob:synth}.}
\label{tab:complexity}
\end{table*}

\section{Proof of Theorem~\ref{thm:milp}}
\label{app:milpproof}

We construct $\Pi(\varphi_S, \KG, P)$, the MILP built by Algorithm~\ref{alg:full}. Let $n = |\VGene|$.

\paragraph{Graph-layer constants.}
Every graph-layer sub-formula $\varphi_G$ of $\varphi_S$ is flat, so its truth
value at gene~$i$ depends only on $\tau(i)$ and $\alpha(i,\cdot)$, which are
fixed data. Let $c_i = [\KG, i \models \varphi_G] \in \bools$, computed by
evaluating Definition~\ref{def:bool} in $O(|\varphi_G|)$ time per gene. No
decision variable is involved.

\paragraph{Variables.}
$x_i \in \bools$ for $i \in \VGene$ ($x_i = 1$ iff $i \in S$);
$z_j \in \bools$ for $j \in \mathcal{J}$; and one $b_\psi \in \bools$ for each
selection-layer sub-formula $\psi$ that is not a negation (a negation
$\neg\psi$ uses $1 - b_\psi$).

\paragraph{Coverage.}
For every patient $j$: $z_j \le \sum_i P_{ij} x_i$, and $z_j \ge x_i$ for every
$i$ with $P_{ij} = 1$. Together these force $z_j = 1$ iff some selected gene is
mutated in patient~$j$, so $\sum_j z_j = m\cdot\mathrm{coverage}(S,P)$. There are
$m + \mathrm{nnz}(P)$ such constraints.

\paragraph{Selection-layer indicators.}
For each $\psi$, the constraints of \textsc{Sel} (Algorithm~\ref{alg:full}):
\begin{itemize}[topsep=2pt,itemsep=1pt]
\item $|S| \ge k$: $\sum_i x_i \ge k\,b_\psi$ and $\sum_i x_i \le k-1+(n-k+1)\,b_\psi$.
  Since $\sum_i x_i$ is an integer in $[0,n]$, these hold exactly when
  $b_\psi = [\sum_i x_i \ge k]$. Other comparators reduce to this one and
  negation (Section~\ref{sec:encoding}).
\item $\mathrm{coverage}(S,P) \ge c$: the same pair with $\sum_j z_j$, $m$, and
  $t = \lceil cm \rceil$ in place of $\sum_i x_i$, $n$, and $k$; exact because
  $\sum_j z_j \ge cm$ iff $\sum_j z_j \ge \lceil cm\rceil$.
\item $\exists g \in S.\,\varphi_G(g)$, with $C = \{i : c_i = 1\}$:
  $b_\psi \ge x_i$ for $i \in C$ and $b_\psi \le \sum_{i \in C} x_i$, which hold
  exactly when $b_\psi = \max_{i\in C} x_i$, i.e.\ some selected gene satisfies
  $\varphi_G$.
\item $\forall g \in S.\,\varphi_G(g)$, with $D = \{i : c_i = 0\}$:
  $b_\psi \le 1 - x_i$ for $i \in D$ and $b_\psi \ge 1 - \sum_{i\in D} x_i$,
  exactly when $b_\psi = 1 - \max_{i\in D} x_i$, i.e.\ no selected gene violates
  $\varphi_G$.
\item $\psi_1 \wedge \psi_2$: $b_\psi \le b_{\psi_1}$, $b_\psi \le b_{\psi_2}$,
  $b_\psi \ge b_{\psi_1} + b_{\psi_2} - 1$, exactly when $b_\psi = b_{\psi_1} b_{\psi_2}$.
\end{itemize}
Finally add $b_{\varphi_S} = 1$ (or $1 - b_{\varphi_S'} = 1$ if
$\varphi_S = \neg\varphi_S'$).

\paragraph{Correctness.}
Fix $S$ and set $x$ to its indicator vector. By the coverage paragraph the
$z_j$ are then determined. By induction on $\psi$, using the exactness noted in
each case, the constraints for the sub-formulas of $\psi$ are satisfiable, and
in every solution $b_\psi = 1$ iff $(\KG,P,S) \models \psi$; the case $\neg\psi$
follows because $1 - b_\psi$ is then exactly the truth value of $\neg\psi$.
Hence $\Pi$ with $x$ fixed is feasible iff $(\KG,P,S) \models \varphi_S$. The
objective $\frac1m\sum_j z_j$ equals $\mathrm{coverage}(S,P)$ in every feasible
solution, so maximizing it solves the zero-operation synthesis problem.

\paragraph{Size.}
Binary variables: $n + m$ plus at most $|\varphi_S|$ indicators. Constraints:
$m + \mathrm{nnz}(P)$ for coverage, $O(n)$ per $\exists/\forall$ quantifier,
and $O(1)$ for every other sub-formula, for a total of
$O(q\cdot n + \mathrm{nnz}(P) + |\varphi_S|)$. Computing the constants takes
$O(|\varphi_S|\cdot n)$ time. \hfill$\square$

\section{Proof of Theorem~\ref{thm:milp-full}}
\label{app:milpproof-full}

\paragraph{Case (i): fixed $\KG$.}
By induction on sub-formulas, Algorithm~\ref{alg:precompile} sets
$\mathit{val}[\psi][v] = [\KG, v \models \psi]$ for every sub-formula $\psi$ and
node $v$: each case of the algorithm is the corresponding clause of
Definition~\ref{def:bool}, and the global modalities are the same at every node.
The cost is $O(|V|)$ per non-modal sub-formula and $O(|V| + |E|)$ per
$\langle\ell\rangle$ or $[\ell]$ sub-formula, so $O(|\varphi_S|\cdot(|V|+|E|))$
in total. The resulting constants $c_i = \mathit{val}[\varphi_G][i]$ play
exactly the role of the flat constants in Appendix~\ref{app:milpproof}, whose
construction and correctness argument apply unchanged. \hfill$\square$ (case i)

\paragraph{Case (ii): symbolic $\KG$ under structural operations.}
Let the permitted operations of Problem~\ref{prob:synth} be given. We encode
the edited graph $\KG'$ directly; only $\KG'$ and the number of operations
matter.
\begin{itemize}[topsep=2pt,itemsep=1pt]
\item \emph{Nodes.} For each node $u$ that may be deleted, $a_u \in \bools$
  ($a_u = 1$ iff $u$ remains); $a_u = 1$ is a constant for every other node.
  For genes, $x_i \le a_i$.
\item \emph{Edges.} For each permitted AddEdge or RemoveEdge on $(u,\ell,u')$,
  $e_{u,\ell,u'} \in \bools$ is the edge's presence in $\KG'$; every other edge
  has its current presence as a constant. The edge is effective when
  $p_{u,\ell,u'} = e_{u,\ell,u'} \wedge a_u \wedge a_{u'}$ (an AND gate, below).
\item \emph{Attributes.} For each permitted Attenuate or Agonize on $(u,a)$, a
  binary $o_{u,a}$ (the operation is used) and a continuous
  $\delta_{u,a} \in [0, \delta_{\max}]$ with $\delta_{u,a} \le \delta_{\max}\,o_{u,a}$;
  then $\alpha'(u,a) = \alpha(u,a) \mp \delta_{u,a}$ is affine. A numeric atom
  $\alpha'(u,a) \ge c$ gets an exact indicator $w$ through
  $\alpha'(u,a) \ge c - M(1-w)$ and $\alpha'(u,a) \le c - \varepsilon + M w$,
  with $M$ bounded by the attribute's range plus $\delta_{\max}$; other
  comparators are analogous, and strict ones use the tolerance $\varepsilon$
  of the theorem statement. Atoms on attributes no operation touches remain
  constants.
\item \emph{Budget.} The number of operations used is
  $\sum (1-a_u) + \sum |e_{u,\ell,u'} - e^0_{u,\ell,u'}| + \sum o_{u,a}$, where
  $e^0$ is the current presence; each $|e - e^0|$ is $e$ or $1-e$ because
  $e^0$ is a constant. Add the constraint that this sum is at most $K$.
\end{itemize}
\emph{Gates.} For binaries $y_1,\ldots,y_r$, an exact AND gate $w = \bigwedge_k y_k$
is $w \le y_k$ for all $k$ and $w \ge \sum_k y_k - (r-1)$; an exact OR gate
$w = \bigvee_k y_k$ is $w \ge y_k$ for all $k$ and $w \le \sum_k y_k$; and
$\neg y$ is $1 - y$. For every graph-layer sub-formula $\psi$ and node $u$,
introduce $w_{\psi,u}$ and tie it to $\psi$ at $u$ in $\KG'$:
\begin{itemize}[topsep=2pt,itemsep=1pt]
\item atoms: the constant or the exact indicator above (a type atom is a
  constant, since no operation changes types);
\item $\wedge,\vee,\neg$: AND, OR, and $1-w$ of the children at $u$;
\item $\langle\ell\rangle\psi'$ at $u$: the OR, over every $u'$ such that
  $(u,\ell,u')$ is a current or permitted edge, of $r_{u,u'} = p_{u,\ell,u'}
  \wedge w_{\psi',u'}$; $[\ell]\psi'$ is $\neg\langle\ell\rangle\neg\psi'$;
\item $\mathsf{E}\,\psi'$: a single variable, the OR over $u' \in V$ of
  $a_{u'} \wedge w_{\psi',u'}$, shared by all $u$; $\mathsf{A}\,\psi'$ is
  $\neg\mathsf{E}\,\neg\psi'$.
\end{itemize}
The selection layer is compiled as in Appendix~\ref{app:milpproof}, except that
$\exists g\in S.\varphi_G$ and $\forall g\in S.\varphi_G$ now read the variable
$w_{\varphi_G,i}$ instead of the constant $c_i$: introduce
$y_i = x_i \wedge w_{\varphi_G,i}$ (an AND gate) and take
$b = \bigvee_i y_i$ for $\exists$, and the dual for $\forall$.

\emph{Correctness.} Fix a feasible assignment of the operation variables; it
determines $\KG'$, and conversely every $(S,\mathcal{O})$ within the budget
determines such an assignment. Because every gate and atom indicator is exact,
induction over sub-formulas shows $w_{\psi,u} = [\KG', u \models \psi]$ for
every $\psi$ and $u$ (deleted nodes are excluded from successor sets and from
$\mathsf{E},\mathsf{A}$ through $a_{u'}$), and then
$b_\psi = [(\KG',P,S) \models \psi]$ as in Appendix~\ref{app:milpproof}.

\emph{Size.} There are $O(|\varphi_S|\cdot|V|)$ variables $w_{\psi,u}$. Each modal
sub-formula adds at most one $r_{u,u'}$ per (current or permitted) successor
pair, at most $|V|^2$, and each such gate has $O(1)$ constraints; each
$\mathsf{E}$ or $\mathsf{A}$ adds $O(|V|)$. The operation variables and the
budget add $O(N_{\mathrm{op}})$, and the selection layer adds
$O(q\cdot|\VGene| + \mathrm{nnz}(P) + |\varphi_S|)$. In total,
$O(|\varphi_S|\cdot|V|^2 + |\VGene| + m + N_{\mathrm{op}})$ variables and
$O(|\varphi_S|\cdot|V|^2 + \mathrm{nnz}(P) + N_{\mathrm{op}})$ constraints.
\hfill$\square$ (case ii)

\section{Proofs of Propositions~\ref{prop:soundness} and~\ref{prop:decidable}}
\label{app:props}

\begin{proof}[Proof of Proposition~\ref{prop:soundness}]
By simultaneous induction on $\varphi$ of both claims; both are needed,
because negation exchanges them.
\emph{Comparisons.} For $\theta\in\{\ge,>\}$, $\rob = x-c > 0$ gives $x>c$, so
$x\,\theta\,c$ holds; $\rob<0$ gives $x<c$, so it fails. $\{\le,<\}$ is
symmetric. For $=$, $\rob = -|x-c|$ is never positive, and $\rob<0$ gives
$x\ne c$.
\emph{Saturated atoms.} $\rob=+1$ exactly when the atom holds, $-1$ exactly
when it fails.
\emph{Negation.} $\rob(\neg\varphi)>0$ iff $\rob(\varphi)<0$, so $\varphi$ fails
by the second claim for $\varphi$ and $\neg\varphi$ holds; the case
$\rob(\neg\varphi)<0$ is symmetric, using the first claim.
\emph{Minimum} ($\wedge$, $[\ell]$, $\mathsf{A}$, $\forall g\in S$). A positive
minimum makes every operand positive, so every operand holds by induction and
the conjunction holds; this includes the empty case, $+\infty$, where the
conjunction is vacuously true. A negative minimum has a negative operand,
which fails by induction, so the conjunction fails.
\emph{Maximum} ($\vee$, $\langle\ell\rangle$, $\mathsf{E}$, $\exists g\in S$)
is dual; the empty case gives $-\infty$, where the disjunction is false.
\end{proof}

\begin{proof}[Proof sketch of Proposition~\ref{prop:decidable}]
Replace each atom of $\varphi_G$ by a fresh proposition and let $\chi$ be the
propositional formula stating that the atoms true at one node are jointly
realizable there: exactly one type holds; an attribute atom implies a type that
carries the attribute; 
for each numeric attribute $a$, let $d_1 < \dots < d_r$ be the constants compared with $a$ in $\varphi_G$ (adding $0$ and $1$ if $a$ is Boolean). They divide the domain of $a$ into the points $d_h$ and the open intervals between and beyond them, and every comparison $\alpha(v,a)\,\theta\, d_h$ has a constant truth value on each of these regions. Introduce one fresh letter per region that meets the domain of $a$ ($\mathbb{R}$, or $\{0,1\}$), require that exactly one region letter for $a$ holds if some type letter whose type carries $a$ holds, and none otherwise, and make each comparison letter on $a$ equivalent to the disjunction of the regions on which the comparison is true. This takes $O(r^2)$ clauses per attribute, so $\chi$ has polynomial size;
and at most one
string equality holds per string attribute. Then $\varphi_G$ is satisfiable over
knowledge graphs iff $\varphi_G \wedge \mathsf{A}\,\chi$ is satisfiable over
multi-relational Kripke structures: a knowledge graph is such a structure, and
conversely each state of a model of $\mathsf{A}\,\chi$ can be given a type and
attribute values realizing its propositions, with edges read off $R_\ell$.
$\mathbf{K}_n$ with the universal modality has the finite model
property~\cite{GorankoPassy92}, so the knowledge graph can be taken finite.
Finally, satisfiability of $\mathbf{K}_n$ with global modalities is in
\textsc{ExpTime}: guess the truth value of each $\mathsf{E}$-subformula (there
are linearly many, and each is constant across states), replace it by that
constant, and check consistency of the resulting description-logic knowledge
base in $\mathcal{ALC}$ with general concept inclusions, which is
\textsc{ExpTime}-complete~\cite{Baader03}. The unimodal special case is already
\textsc{ExpTime}-hard~\cite{Hemaspaandra96}.
\end{proof}

\section{The Kripke embedding}
\label{app:kripke}\label{app:kripke-instance}

Verification and synthesis (Section~\ref{sec:verify}) use
Definition~\ref{def:bool} directly. This subsection connects the graph layer
to standard modal logic, which is what lets us settle its decidability
(Proposition~\ref{prop:decidable}). A classical Kripke structure
$M = (W, R, \AP, \lambda)$ has states~$W$, a transition relation~$R$, atomic
propositions~$\AP$, and a labeling function $\lambda : W \to 2^{\AP}$ mapping
each state to the set of propositions it satisfies. Temporal and modal logics
are interpreted over such structures.

\begin{definition}[Multi-relational Kripke Embedding]\label{def:kripke}
The KG $\KG = (V,E,\tau,\alpha)$ induces a multi-relational Kripke structure
\[
M_\KG \;=\; \bigl(W,\, \{R_\ell\}_{\ell \in L},\, \AP,\, \lambda\bigr)
\]
where:
\begin{itemize}[topsep=2pt,itemsep=1pt]
  \item $W = V$ (states are nodes);
  \item $R_\ell = \{(u,v) : (u,\ell,v) \in E\}$ for each $\ell \in L$
        (one accessibility relation per edge label);
  \item $\AP = \{\tau(v)=t : t \in T\} \;\cup\;
        \{\alpha(v,a)\,\theta\, c : a \in \mathrm{NumAttr},\;
        \theta \in \{<,\le,=,\ge,>\},\; c \in \mathbb{Q}\} \;\cup\;
        \{\alpha(v,a')=s : a' \in \mathrm{StrAttr},\; s \in \Sigma^*\}$
        (the numeric/string attribute-name split is spelled out in
        Remark~\ref{rem:attrnames-split});
  \item $\lambda(v) = \{\varphi \in \AP : \KG, v \models \varphi\}$
        (atomic propositions satisfied at node~$v$).
\end{itemize}
\end{definition}

Reading $\langle\ell\rangle$ and $[\ell]$ over $R_\ell$, and $\mathsf{E}$,
$\mathsf{A}$ as the global modalities ranging over all states, the standard
Kripke semantics agrees with Definition~\ref{def:bool} by construction:
$M_\KG, v \models \varphi_G$ if and only if $\KG, v \models \varphi_G$, for every
node~$v$ and graph-layer formula~$\varphi_G$. The embedding adds no new meaning;
it identifies the graph layer as a multi-modal logic, so that results about
such logics apply.

\begin{remark}[What the Kripke embedding buys us]
\label{rem:kripke}
The graph layer is therefore the multi-modal logic $\mathbf{K}_n$
($n=|L|$)~\cite{Blackburn01} extended with the global modalities
$\mathsf{E},\mathsf{A}$~\cite{GorankoPassy92}, over atomic propositions that are
attribute comparisons; for example,
$\langle \textsc{ParticipatesIn}\rangle\,(\tau(v)=\textsc{Pathway})$ is a one-edge
adjacency check. This identification makes decidability answerable
(Proposition~\ref{prop:decidable}), and the precompilation pass of
Section~\ref{sec:encoding} is ordinary bottom-up model checking on~$M_\KG$.
\end{remark}

\begin{example}[The Kripke embedding of Example~\ref{ex:nsclc}]
\label{ex:kripke-instance}
For the nine-node subgraph of Figure~\ref{fig:subgraph}
($W = \{\textsc{EGFR},\allowbreak \textsc{KRAS},\allowbreak \textsc{TP53},\allowbreak \textsc{PIK3CA},\allowbreak \textsc{MUC16},\allowbreak
\text{RTK--RAS},\allowbreak \text{PI3K},\allowbreak \text{TP53 pathway},\allowbreak \textsc{NSCLC}\}$, the three pathways abbreviated), the induced
$M_\KG = (W, \{R_\ell\}_{\ell\in L}, \AP, \lambda)$ has:
\begin{align*}
&R_{\textsc{AssociatedWith}} = \bigl\{\, (\textsc{EGFR},\textsc{NSCLC}),\\
&\quad (\textsc{KRAS},\textsc{NSCLC}), (\textsc{PIK3CA},\textsc{NSCLC}), \\
&\quad (\textsc{TP53},\textsc{NSCLC}), (\textsc{MUC16},\textsc{NSCLC}) \,\bigr\}
\end{align*}
\begin{align*}
&R_{\textsc{ParticipatesIn}} = \{(\textsc{EGFR}, \text{RTK--RAS}),\ (\textsc{KRAS}, \text{RTK--RAS}), \\
&\quad (\textsc{PIK3CA}, \text{PI3K}),\ (\textsc{TP53}, \text{TP53 pathway})\}
\end{align*}
and $R_{\textsc{InteractsWith}}$ relates every two distinct genes among
\textsc{EGFR}, \textsc{KRAS}, \textsc{PIK3CA}, and \textsc{TP53}, in both
orientations. Restricted to the propositions relevant to Example~\ref{ex:passenger},
$\lambda(\textsc{KRAS}) \supseteq \{\tau=\textsc{Gene},\ \text{driver}=1,\ \text{essentiality}<-0.5,\ \text{druggable}=1\}$
and $\lambda(\textsc{MUC16}) \supseteq \{\tau=\textsc{Gene},\ \text{driver}=0,\ \text{essentiality}\ge -0.5\}$.
This is small enough to check by hand that
$\langle\textsc{ParticipatesIn}\rangle(\tau(v)=\textsc{Pathway})$ holds at
\textsc{EGFR} (via its RTK--RAS edge) but the passenger predicate
$\varphi_{\mathrm{pass}}$ of Example~\ref{ex:passenger} holds only at \textsc{MUC16}.
\end{example}


\end{document}